\documentclass[%
 aip,
 amsmath,amssymb,
 reprint,%
]{revtex4-1}

\usepackage{graphicx}
\usepackage{dcolumn}
\usepackage{bm}
\usepackage[utf8]{inputenc}
\usepackage[T1]{fontenc}
\usepackage{mathptmx}
\usepackage{etoolbox}
\usepackage{hyperref}

\makeatletter
\def\@email#1#2{%
 \endgroup
 \patchcmd{\titleblock@produce}
  {\frontmatter@RRAPformat}
  {\frontmatter@RRAPformat{\produce@RRAP{*#1\href{mailto:#2}{#2}}}\frontmatter@RRAPformat}
  {}{}
}%
\makeatother

 \newcommand{\pid}{-\Pi_{ij}D_{ij}}

 \newcommand{\m}[1]{\mathbf #1} 
 \newcommand{\jde}{\m{j}\cdot \m{E}}

\begin{document}

\preprint{AIP/123-QED}

\title[Dissipation in Reconnection]{Energy Dissipation in Turbulent Reconnection}
\author{R. Bandyopadhyay}%
\email{riddhib@princeton.edu}
 \affiliation{Department of Astrophysical Sciences, Princeton University, Princeton, NJ 08544, USA}

\author{A. Chasapis}
\affiliation{Laboratory for Atmospheric and Space Physics, University of Colorado Boulder, Boulder, CO 80309, USA} 
 
\author{W.~H. Matthaeus}
\affiliation{Department of Physics and Astronomy, University of Delaware, Newark, DE 19716, USA}
\affiliation{Bartol Research Institute, University of Delaware, Newark, DE 19716, USA}	

\author{T.~N. Parashar}
\affiliation{School of Chemical and Physical Sciences, Victoria University of Wellington, Wellington 6012, New Zealand}

\author{C.~C. Haggerty}
\affiliation{University of Hawaii, Institute for Astronomy, Honolulu 96822, USA}

\author{M.~A. Shay}
\affiliation{Department of Physics and Astronomy, University of Delaware, Newark, DE 19716, USA}
\affiliation{Bartol Research Institute, University of Delaware, Newark, DE 19716, USA}	

\author{D.~J. Gershman}
\affiliation{NASA Goddard Space Flight Center, Greenbelt, Maryland 20771, USA}

\author{B.~L. Giles}
\affiliation{NASA Goddard Space Flight Center, Greenbelt, Maryland 20771, USA}

\author{J.~L. Burch}
\affiliation{Southwest Research Institute, San Antonio, Texas 78238, USA}

\date{\today}

\begin{abstract}
We study the nature of pressure-strain interaction at reconnection sites, detected by NASA's Magnetospheric Multiscale (MMS) Mission. We employ data from a series of published case studies, including a large-scale reconnection event at the magnetopause, three small-scale reconnection events at the magnetosheath current sheets, and one example of the recently discovered electron-only reconnection. In all instances, we find that the pressure-strain shows signature of conversion into (or from) internal energy at the reconnection site. The electron heating rate is larger than the ion heating rate and the compressive heating is dominant over the incompressive heating rate in all cases considered. The magnitude of thermal energy conversion rate is close to the  electromagnetic energy conversion rate in the reconnection region. Although in most cases the pressure-strain interaction indicates that the particle internal energy is increasing, in one case the internal energy is decreasing. These observations indicate that the pressure-strain interaction can be used as an independent measure of energy conversion and dynamics in reconnection regions, in particular independent of measures based on the electromagnetic work. Finally, we explore a selected reconnection site in a turbulent Particle-in-Cell (PIC) simulation which further supports the observational results.
\end{abstract}

\maketitle

\section{\label{sec:intro}Introduction}
Magnetic reconnection is a fundamental process of energy conversion in plasmas. During reconnection, magnetic energy is converted into particle energy. Energy dissipation during reconnection has been a topic of increasing interest because of its potentially significant 
effects on the plasma. This is of central interest in several problems of space physics, such as understanding the heating of the solar corona and solar wind, and acceleration of energetic particles~\cite{Montgomery1980RNPTES, Bruno2005LRSP, Matthaeus2011SSR, Verscharen2019LRSP}. Despite its importance and compelling nature, the nature of energy conversion in magnetic reconnection in weakly-collisional plasmas remains incompletely understood. Several questions remain such as what fraction of energy is converted to bulk velocity of the particles and what fraction of energy is dissipated to the internal energy of each charged particle species? Several previous works, using data from experiments and simulations, have addressed this point (see \cite{Yamada2015PoP_energy_partition, Yoo2017JGR_electron_heating, Fadanelli2021JGR_energy_conversion} and references therein.)

 
There have been various efforts to identify the specific \textit{mechanisms} involved in collisionless dissipation. Several heating mechanisms such as Landau/transit-time resonances~\cite{Leamon1999JGR, Howes2008JGR}, stochastic heating~\cite{Chandran2010ApJ}, cyclotron resonances~\cite{Hollweg2002JGR_cyclotron_resonance}, kinetic instabilities~\cite{Gary1993Book}, and magnetic pumping~\cite{Lichko2017ApJ_magnetic_pumping} have been proposed as contributing candidates. Another approach, followed in this paper, focuses more broadly on the general discussion of the  pathways leading to dissipation: transfer across scales, transfer between particles and electromagnetic fields, and transfer between flow energies and internal energies for each species.

This approach
\cite{Yang2017PoP,Yang2017PRE} eliminates the simplifying assumptions that accompany the identification of specific dominant mechanisms. All available channels, apart from collisions, are formulated based on the multi-species Vlasov equation. Therefore, an identification of a dominant pathway to dissipation includes all kinetic processes that lead to dissipation, and consequently, is more general and broadly applicable.

\section{Energy conversion channels}\label{definitions}
For a collisionless plasma consisting of charged  
species labeled by $\alpha$, the total 
energy density at each point $\mathbf{x}$ at a given time $t$ 
can be separated into 
the electromagnetic energy density,
\begin{eqnarray}
\mathcal{E}^m({\bf x},t)=
{\frac{1}{8\pi}}\left({\bf B}^2({\bf x},t)
+{\bf E}^2({\bf x},t)\right)
\end{eqnarray}
and the sum over species of the individual 
particle
kinetic energy density
\begin{eqnarray}
\mathcal{E}_{\alpha} =
{\frac{1}{2}} m_\alpha \int{ |{\bf{v}}|^2 
f_\alpha \left({\bf x},{\bf v},t\right) d{\bf v}}.
\end{eqnarray}
Here 
${\bf B}$ and ${\bf E}$ are magnetic and electric fields, respectively, 
$m_\alpha$ is the mass of particles of species $\alpha$,
and $f_\alpha$ is the velocity distribution function 
of particles of type $\alpha$, varying in position and time. 
The collective motion is quantified by the 
fluid velocity ${\bf u_\alpha}$ defined by 
$n_\alpha {\bf u}_\alpha = \int {\bf v} f_\alpha d{\bf{v}}$,
where $n_\alpha = \int f_\alpha d{\bf v}$ is the number density of species $\alpha$.
Hereafter we refer to these energy densities, for brevity, 
simply as energies.

Separating 
the kinetic energy 
into average and random parts facilitates
the understanding the energy conversion processes.
The fluid flow kinetic energy of species $\alpha$ is 
\begin{eqnarray}
\mathcal{E}^f_\alpha={\frac{1}{2}}n_\alpha m_\alpha |\bf{u}_\alpha|^2
\end{eqnarray}
and the corresponding thermal (or internal, or random) energy is
\begin{eqnarray}
\mathcal{E}^{th}_\alpha={\frac{1}{2}} m_\alpha \int{\left(\bf{v}-\bf{u}_\alpha\right)^2 f_\alpha \left({\bf x},{\bf v},t\right) d\bf{v}},
\end{eqnarray}
we have $\mathcal{E}_{\alpha}=\mathcal{E}^f_\alpha+\mathcal{E}^{th}_\alpha$.

The time evolution of the energies follows directly from manipulations of the Vlasov equation and 
Maxwell's equations. 
The fluid flow energy evolves in time 
according to 
\begin{eqnarray}
\partial_t \mathcal{E}^{f}_\alpha + \nabla \cdot \left( \mathcal{E}^{f}_\alpha \bf{u}_\alpha \right) +
\nabla \cdot \left( \bf{P}_\alpha \cdot {\bf u}_\alpha \right) && \nonumber \\ = 
\left( \bf{P}_\alpha \cdot \nabla \right) \cdot {\bf u}_\alpha+ 
n_\alpha q_\alpha {\bf u}_\alpha  \cdot  \bf{E} .
\label{eq:Ef}
\end{eqnarray}
Similarly 
one obtains the 
time evolution equation for the internal energy \footnote{Some readers may prefer to use terminology such as 
``thermal energy'' or ``random kinetic energy'' to refer to the 
quantity $\mathcal{E}^{th}_\alpha$.} of species $\alpha$, 
\begin{eqnarray}
\partial_t \mathcal{E}^{th}_\alpha + \nabla \cdot \left( \mathcal{E}^{th}_\alpha \bf{u}_\alpha \right) + \nabla \cdot \bf{h}_\alpha &=& -\left( \bf{P}_\alpha \cdot \nabla \right) \cdot \bf{u}_\alpha .
\label{eq:Et}
\end{eqnarray}

where $\bf{h}$ is the heat flux vector. 

Finally, using 
the Maxwell's equations,
the equation governing $\mathcal{E}^m$
can be written as
\begin{eqnarray}
\partial_t \mathcal{E}^{m} + {\frac{c}{4\pi}} \nabla \cdot \left( \bf{E} \times \bf{B} \right) &=& -\bf{j} \cdot \bf{E}
\label{eq:Em}
\end{eqnarray}
where $\bf{j}=\sum_{\alpha} \bf{j}_\alpha$ is the total electric current density, and
${\bf j}_\alpha=n_\alpha q_\alpha \bf{u}_\alpha$ is the electric current density of species $\alpha$.

Several features of energy transfer can be seen from Eqs.~(\ref{eq:Ef}), (\ref{eq:Et}) and (\ref{eq:Em}). 
First, all the divergence terms on the left-hand sides are
transport terms that do not change the total amount of energy of the respective 
types, but simply move energy from one location to another. 
These transport 
terms integrate to zero for suitable boundary conditions and may be  important in reconciling the energy balance at any point in space and time~\cite{Eastwood2020PRL_flux}. 
However, here we are not concerned with transport effects
as the emphasis is on 
quantifying conversion between different types of 
energy\footnote{We do not suggest that transport effects such as
heat flux and convective heat transport are small, as in many circumstances
these are significant or even dominant contributions to the 
balance of equation \ref{eq:Et}; however these terms do not
exchange energy between different forms, and it is the exchange between 
different pathways or channels that is our main interest here.}.
Therefore, we will only discuss the 
terms on the right-hand sides of these equations that
are responsible for 
conversion of energy from one form to another. 

Examining these terms,
it is evident that the $\bf{j}_\alpha \cdot {\bf E}$ terms 
convert electromagnetic energy into or out of flow energy for each species $\alpha$. 
All changes of internal energy of each species 
are accomplished solely by the pressure-stress interaction, 
$\left( {\bf P}_\alpha \cdot \nabla \right) \cdot {\bf u}_\alpha 
= P^{(\alpha)}_{ij} \nabla_i\, u^{(\alpha)}_j$.
      
Quantities 
such as $\bf{j} \cdot {\bf E}$
and $\left( {\bf P} \cdot \nabla \right) \cdot {\bf u}$ are not single signed, as
energy may be transferred into, or out of, 
the electromagnetic fields, and likewise, into 
or out of the collective fluid motion
of each species $\alpha$. 
While the distributions of these quantities
are not sign-definite, the expectation is
that when there is 
net dissipation and heating, the
appropriate sign indicating net transfer into random motions 
will be favored. This has been seen in magnetosheath 
observations
\cite{Bandyopadhyay2020bPRL} and in plasma 
simulations in decaying turbulence \cite{Yang2018MNRAS}. 
Therefore, with some care, these quantities may be used 
to trace the flow of energy through different channels leading
to dissipation. 
 
Further decomposition of the pressure-strain is convenient. 
A standard procedure for decomposing the pressure
tensor $P^{(\alpha)}_{ij}$ and 
the stress tensor  $S^{(\alpha)}_{ij}=\nabla_i u^{(\alpha)}_j$
is to separate out the trace.
One then defines 
$P^{(\alpha)}_{ij} = p_\alpha \delta_{ij}  + \Pi^{(\alpha)}_{ij}$
where $p_\alpha = \frac13 P^{(\alpha)}_{jj}$ and , $\Pi_{ij} = P_{ij} - p\delta_{ij}$.
Similarly the stress tensor  
is conveniently decomposed
as $S^{(\alpha)}_{ij}  =
\frac13\theta_\alpha\delta_{ij} + D^{(\alpha)}_{ij} + \Omega^{(\alpha)}_{ij}$
where $\theta=\nabla \cdot {\mathbf u}$, $D^{(\alpha)}_{ij} = \frac12\left ( \nabla_i u^{(\alpha)} _j + \nabla_ju^{(\alpha)}_i\right ) $,
and $\Omega^{(\alpha)}_{ij} = \frac12\left ( \nabla_i u^{(\alpha)} _j - \nabla_ju^{(\alpha)}_i \right ) $. 
Then 
we see that the pressure stress interaction separates as 
\begin{eqnarray}
\left( \bf{P}_\alpha \cdot \nabla \right) \cdot {\bf u}_\alpha = p^{(\alpha)} \theta^{(\alpha)} + \Pi^{(\alpha)}_{ij}D^{(\alpha)}_{ij} \label{eq:decomp}.
\end{eqnarray}

Below, we study quantities 
that participate in 
termination of the inertial-range cascade, 
while opening 
channels for production of velocity space structures (higher-order moments)~\cite{Servidio2017PRL, Schekochihin2016JPP} that lead eventually to collisional effects.  
In this sense the 
${\bf j     } \cdot {\bf E}$ and $\left( {\bf P} \cdot \nabla \right) \cdot {\bf u}$ transfer channels 
provide the gateways 
between fluid scale effects and 
kinetic dissipation. 

Based on this discussion, there are at least two key motivations for studying the pressure-strain interaction:
Changes in the random energy, $\mathcal{E}^{th}_{\alpha}$, take place only through the pressure-strain term~\footnote{We emphasize again that the transport terms such as heat flux can change the local value of $\mathcal{E}^{th}_{\alpha}$ too, but here we are concerned only with \textit{conversion} of energy.} (Eq.~\ref{eq:Et}); and the pressure-strain interaction advances the ohmic energy dissipation $\jde$, in that this approach tracks the electron versus proton heating, as well as compressive $(-p \theta)$ versus incompressive $(-\Pi_{ij}D_{ij})$ heating.

However, we note that an evaluation of $\left( {\bf P} \cdot \nabla \right) \cdot {\bf u}$ requires accurate measurement of the full pressure tensor as well as computation of spatial derivatives down to kinetic scales~\citep{Dunlop1988ASR, Paschmann1998ISSI}. This requires the availability of multi-point measurement with very high resolution, which was not possible before the Magnetospheric Multiscale (MMS) Mission~\citep{Burch2016SSR}. The MMS data set provides the first opportunity to study these quantities using observational data. 

The dissipation through the pressure-strain interaction is agnostic of dissipation mechanism.
Although several kinetic processes can participate in the energy dissipation process, here we
only investigate one of them: magnetic reconnection, which  plays a central role in the dynamics of Earth's magnetosphere~\cite{Sonnerup1979Book, Sonnerup2013book}.

\section{MMS Observations}\label{sec:mms}
MMS data provide the dual advantages of high-time cadence and simultaneous multi-spacecraft measurements at very small inter-spacecraft separations. This combination enables the study of the nature of kinetic plasma at small scales with an unprecedented level of accuracy and resolution. In this study, we use magnetic field $(\mathbf{B})$ data from the Fluxgate Magnetometers (FGM)~\cite{Russell2016SSR}, electric field $(\mathbf{E})$ data from the Electric Field Double Probes (EDP)~\cite{Ergun2016SSR_EDP_MMS, Lindqvist2016SSR_SPDP_MMS}, and ion and electron densities $(n_{\mathrm{i}}, n_{\mathrm{e}})$, bulk velocities $(\mathbf{u}_{\mathrm{i}}, \mathbf{u}_{\mathrm{e}})$, and pressure tensor $(\mathbf{P}_{\mathrm{i}}, \mathbf{P}_{\mathrm{e}})$ from the Fast Plasma Investigation (FPI)~\cite{Pollock2016SSR}, using the computed moments of the ion and electron distribution functions. We mainly use MMS burst resolution data which provide magnetic-field measurements (FGM) at $128$ samples/second,  ion density and velocity (FPI-DIS) at $150$ ms, electron density and velocity (FPI-DES) at $30$ ms, and electric field data at $8192$ samples/second. For the electron-only reconnection case (see Section~\ref{sec:phan}) we use the FPI quarter-moments data product, which generates $7.5$ ms electron and $37.5$ ms ion data products by separating the individual energy sweeps~\cite{Rager2018GRL_quarter_moment}. We employ a multi-spacecraft technique, similar to the curlometer technique~\cite{Dunlop1988ASR, Paschmann1998ISSI}, enabling evaluation of the velocity strain tensor. We average the pressure tensor over the four MMS spacecraft. To calculate the Zenitani measure~\cite{Zenitani2011PRL} $\jde^{\prime}$, where $\mathbf{E}^{\prime} = \mathbf{E} + \mathbf{V}_{\mathrm{e}} \times \mathbf{B}$, we use particle density and velocity, $\mathbf{j} = q n_{\mathrm{e}} (\mathbf{V}_{\mathrm{i}} - \mathbf{V}_{\mathrm{e}})$ ($q$ is the electronic charge) and average over the 4 spacecraft.

\section{\label{sec:burch}Large-Scale Reconnection}
Magnetic reconnection at the large scale, such as those in the Earth’s magnetopause or magnetotail, can act as a ``driver", injecting energy into the system and drive turbulence~\cite{Pucci2017ApJ}. Magnetopause reconnection is understood to be an important process for coupling solar wind mass and momentum to the Earth's magnetosphere. Since its launch in 2015, MMS has observed 
numerous 
reconnection
events along the sunward boundary of Earth's magnetosphere where the interplanetary magnetic field reconnects with the terrestrial magnetic field~\cite{Burch2016Science, Burch2016GRL, Chen2016GRL}. Here, we first investigate the nature of the
pressure-strain interaction at an MMS-identified magnetopause reconnection site to gain insights into the nature of dissipation during the large-scale reconnection process at the magnetopause. 

We begin with an example of the MMS measurements of the reconnection diffusion region
around an X-line within the magnetopause current sheet, reported in \citet{Burch2020GRL}. On 15 April 2018, the MMS tetrahedron encountered a reconnection site on the dayside magnetopause and observed the conversion of magnetic energy to particle energy. The details  are already given in \citet{Burch2020GRL}, so here we only focus on the dissipation measures at the reconnection site. 

\begin{figure}
\includegraphics[width=\linewidth]{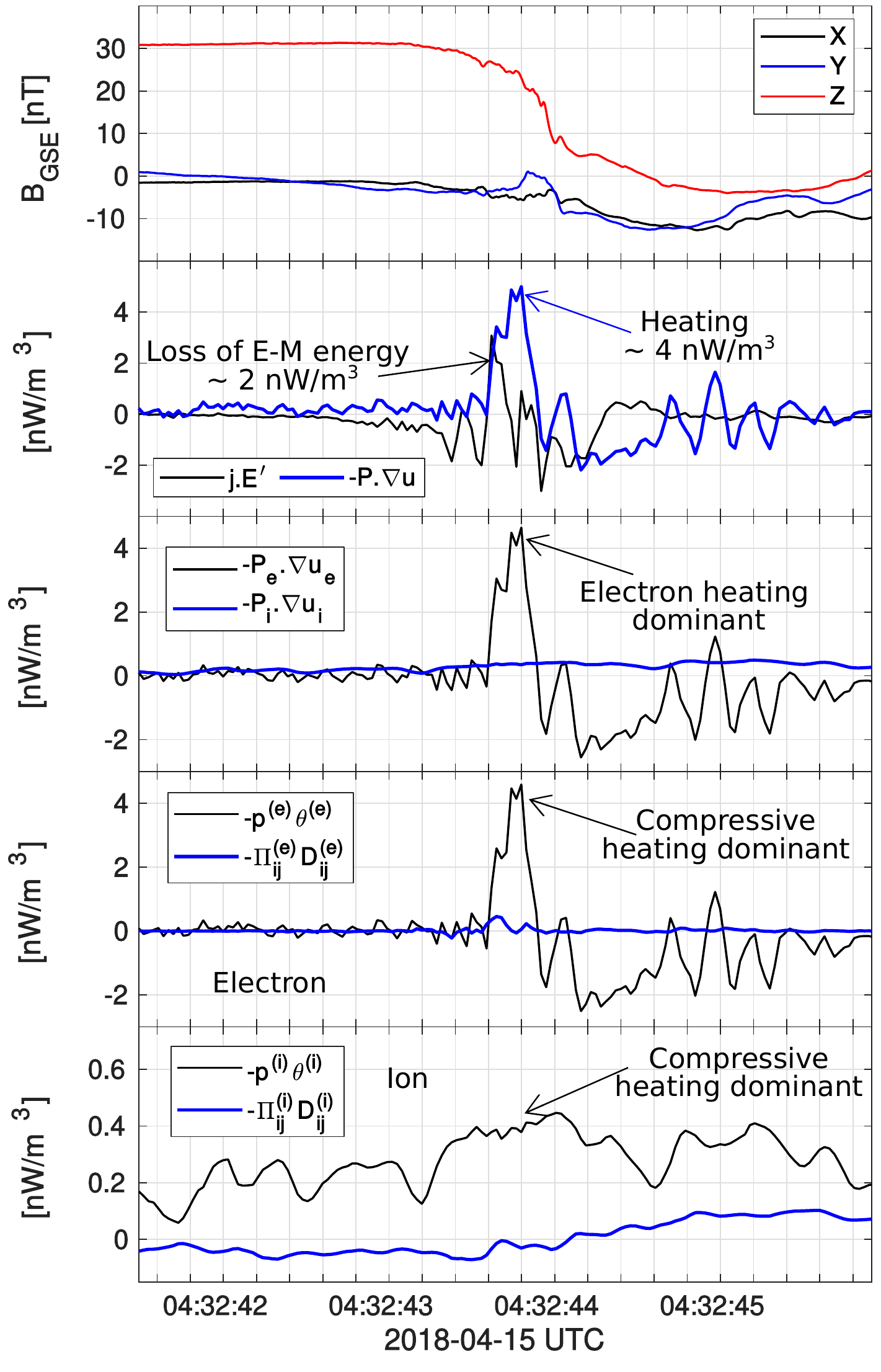}
\caption{\label{fig:burch}Magnetopause reconnection from \citet{Burch2020GRL}.}
\end{figure}

The top panel of Fig.~\ref{fig:burch} plots the magnetic field components in geocentric solar ecliptic (GSE) coordinates~\cite{Franz2002PSS}, measured by MMS1.
The second panel compares the energy conversion rate from electromagnetic energy, as quantified by the Zenitani measure (in narrow black), and the energy conversion rate to random energy, measured by the total pressure-strain (in broad blue). Both conversion rates show an enhanced signal at the reconnection region. Further, the two quantities are close, but the value of  $\left( {\bf P} \cdot \nabla \right) \cdot {\bf u}$ is somewhat higher than ${\bf J} \cdot {\bf E}^{\prime}$. This observation suggests that the interval energy may come from sources other than the electromagnetic energy, possibly the fluid kinetic energy.

The third panel plots the total pressure-strain rate from electron and ions. This panel shows that the electrons are responsible for the majority of the heating.

The bottom two panels show the pressure-strain interaction decomposed into the compressive and incompressive channels (Eq.~\ref{eq:decomp}), for electrons and ions separately. The two panels show that for both charged species, the compressive channel is responsible for 
most of the energy conversion.

\section{\label{sec:wilder}Small-Scale Reconnection}
\vspace{-0.15in}
Strong current sheets form between the margins of magnetic islands in a turbulent system, and reconnection occurs in these current sheets. Reconnecting X-lines near thin current sheets are an important
element of turbulent cascade~\cite{Carbone1990PoFA_coherent_structure, Servidio2009PRL, Cothran2003GRL_3Dreconnection, Retino2007NaturePh, Comisso2018ApJ_plasmoid_mediated}. In addition to the standard reconnection in large-scale current sheets at the magnetopause, MMS has identified many reconnection events in the magnetosheath current sheets, downstream of the bow shock~\cite{Voros2017JGR, Wilder2017PRL, Wilder2018JGR_reconnection}. Similar to the previous section, here we study the nature of pressure-strain interaction in the MMS-identified reconnection sites in the turbulent magnetosheath to find the incompressive versus compressive conversion of energy and
also ion versus electron heating in turbulent reconnection. We focus on three samples with varying out-of-plane guide magnetic-field strengths, reported previously by \citet{Wilder2018JGR_reconnection}

\subsection{\label{sec:low}Weak Guide Field}
\begin{figure}
\includegraphics[width=\linewidth]{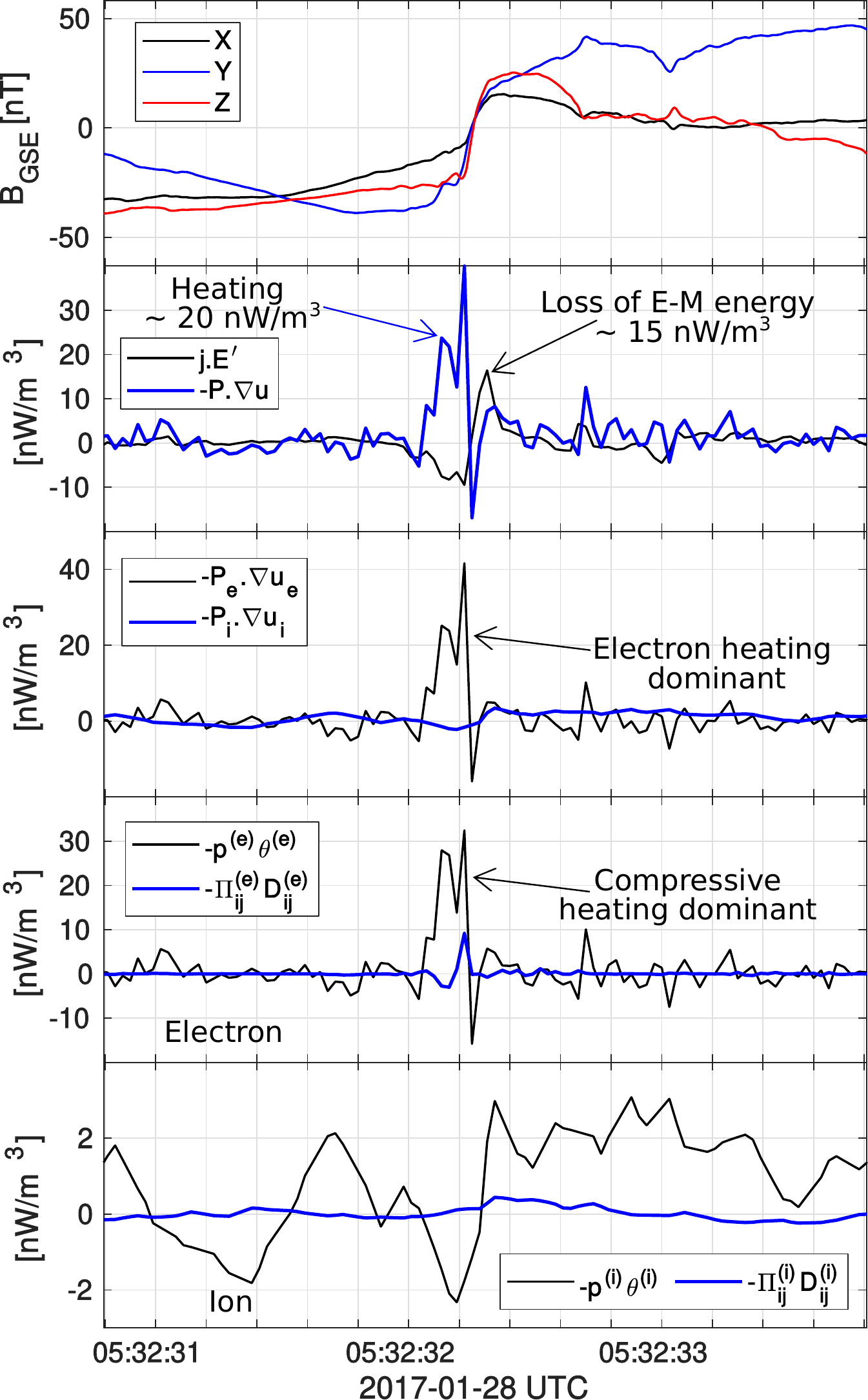}
\caption{\label{fig:small}Magnetosheath reconnection with small guide field, reported by \citet{Wilder2017PRL}.}
\end{figure}

\vspace{-0.1in}
We start with a sample of magnetosheath reconnection in presence of a very small guide field $(B_{\mathrm{g}}\approx 0.1)$. Figure~\ref{fig:small} shows the magnetic field components and energy conversion measures in the same format as the previous section. 

From the second panel, both $\mathbf{j}\mathbf{\cdot}\mathbf{E}^{\prime}$ and $-\mathbf{P} \mathbf{\cdot} \mathbf{\nabla} \mathbf{u}$ are of positive sign and similar magnitude. We interpret this as particles being energized by converting energy from the electromagnetic energy and then almost all of that energy is being dissipated into thermal degree of freedom. The heating rate is more than the rate of electromagnetic energy dissipation, which suggests that the internal energy originates 
from flows as well. We note here that there is a peak of $-\mathbf{P} \mathbf{\cdot} \mathbf{\nabla} \mathbf{u}$ with a value twice as large $(\sim 40 \mathrm{nW}/\mathrm{m}^3)$, very close to the actual magnetic reversal point. However, the large but narrow peak is immediately followed by an almost equally large and narrow peak in the negative direction. Thus, in considering the dissipation rate in the region surrounding the peak in $\mathbf{j}$ that captures the full current structure, the primary contribution  comes from the $(\sim 20 \mathrm{nW}/\mathrm{m}^3)$ peak. Another point worth remarking is that the peaks of the two quantities are slightly shifted in time (space). Although this could be due a physical effect as the pressure strain interaction is shown to be active {\em near} and {\em not in the same location} as the current sheets \cite{Parashar2016ApJ_prop, Yang2017PoP, Bandyopadhyay2020aPRL}. However, the limitations of the curlometer technique and other approximations in data analysis could also be responsible for it.

From the third panel, similar to the previous section, electrons contribute to most of the heating. In fact, ions show a signature of cooling rather than heating. Finally, the fourth panel indicates that majority of the electron heating occurs through the compressive channel. Similarly, the last panel shows that although ions are cooling, most of this effect 
occurs through the compressive channel.

\subsection{\label{sec:moderate}Moderate Guide Field}
\begin{figure}
\includegraphics[width=\linewidth]{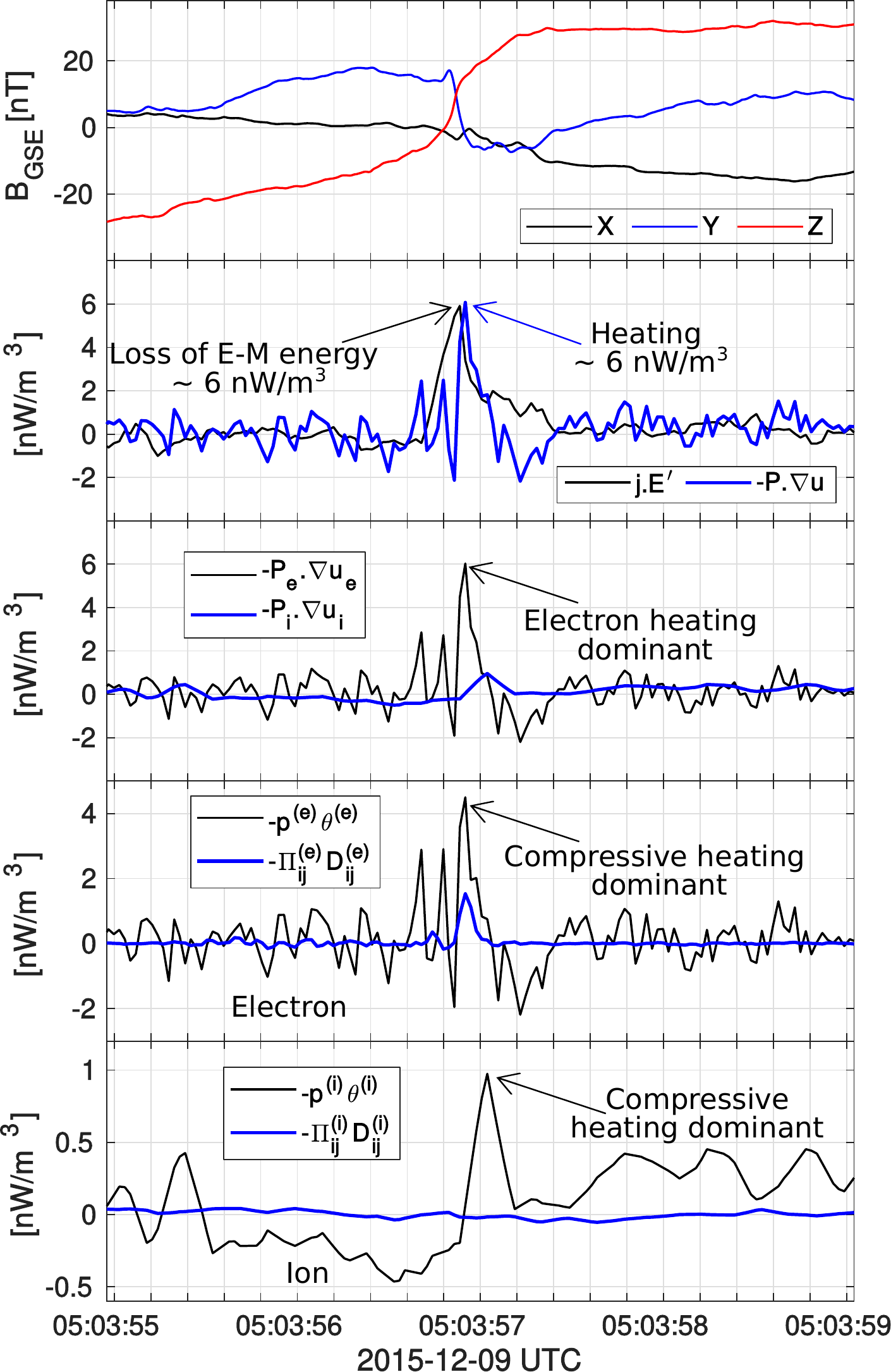}
\caption{\label{fig:moderate}Magnetosheath reconnection with moderate guide field from \citet{Wilder2018JGR_reconnection}.}
\end{figure}

\vspace{-0.1in}
Next, we consider another magnetosheath reconnection with medium guide field strength $(B_{\mathrm{g}}\approx 0.5)$. The MMS observations are shown in Fig.~\ref{fig:moderate}. 
In this case, both electrons and ions show signatures of heating. The total heating rate is almost exactly equal to the rate of electromagnetic energy dissipation rate (second panel), indicating only a small fraction of the magnetic energy, after conversion, 
persists as 
fluid kinetic energy. Again, we find that the electron heating is much stronger than the ion. Further, the compressive channel of heating is much stronger than the incompressive one for both charged species.

\subsection{\label{sec:high}High Guide Field}
\begin{figure}
\includegraphics[width=\linewidth]{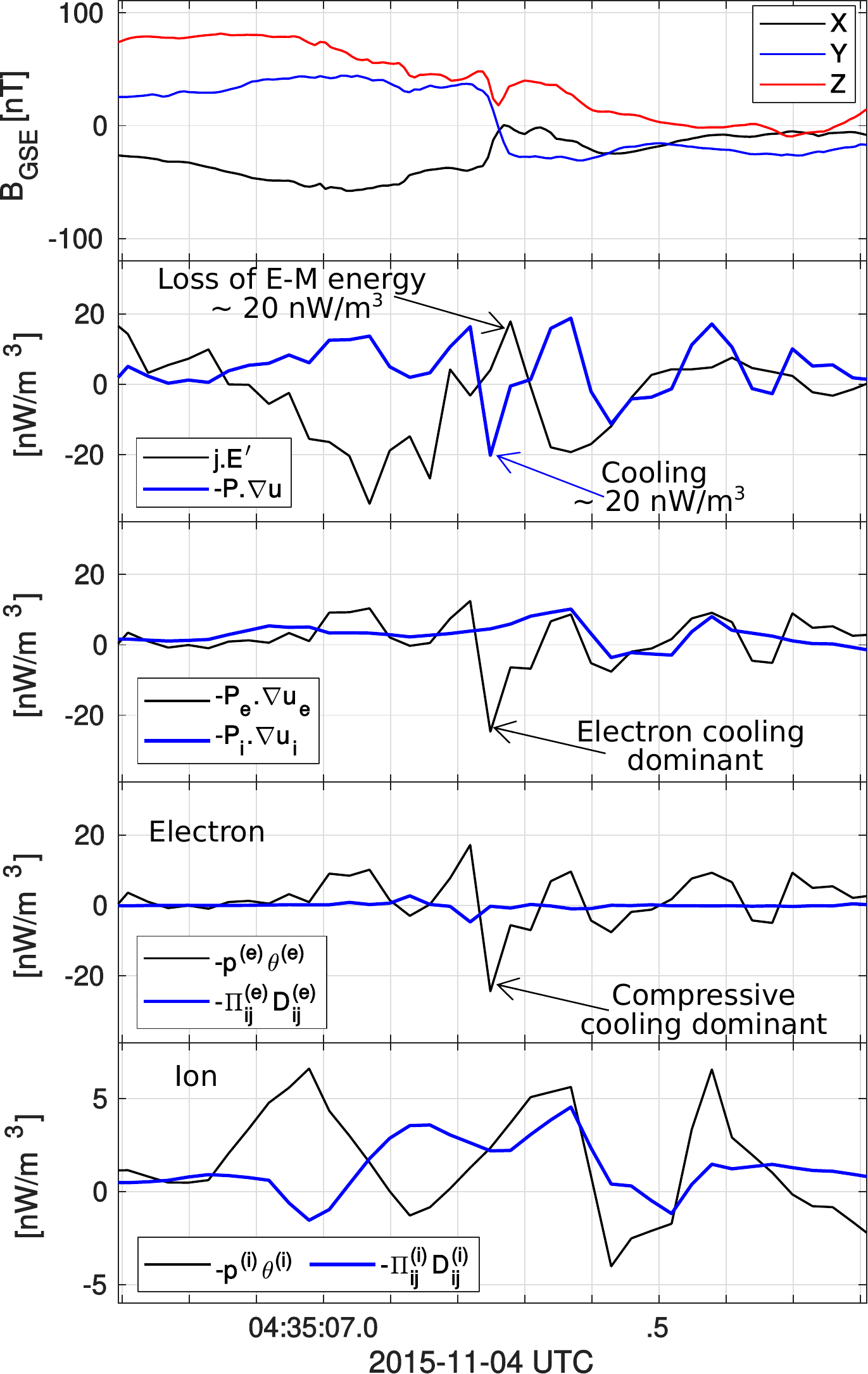}
\caption{\label{fig:high}Magnetosheath reconnection with high guide field~\cite{Wilder2018JGR_reconnection}. The particles are losing internal energy although electromagnetic energy is being dissipated.}
\end{figure}

\vspace{-0.2in}
As a final case of magnetosheath reconnection, we show an example 
with high guide field $(B_{\mathrm{g}}\approx 1.3)$ in Fig.~\ref{fig:high}. 
Here, we see that the electrons are cooling instead of heating, while the ions show a signature of slight heating. The electron cooling is the more prominent effect, and so the rate of change of 
total particle internal energy indicates cooling. 
Although this electron cooling is only seen in this high-guide field case, with this limited number of case studies, it is not possible to conclude whether it is the high guide field strength or some other parameters such as the upstream Alfv\'en flow speed that is responsible for the electron cooling. However, this example shows that the pressure-strain interaction can be used as an independent diagnostic at reconnection sites to see whether particles are heating or cooling. In this case, the electron pressure is enhanced near the reconnection site (see Fig 6 of \citet{Wilder2018JGR_reconnection}), and so the mechanical pressure may possibly cause a local expansion of the plasma. This expansion, in turn, results in local cooling of the plasma, which is manifested through the $p \theta$ term. We note that on either side of the central electron cooling region, there are nearby regions of electron heating, separated by several ion inertial scales and coinciding with the flanks of the central current enhancement (not shown; see \citet{Wilder2018JGR_reconnection}).
It appears in this case that the central electron diffusion region is cooling while the broader ion-scale diffusion region may be increasing its internal energy, resulting in overall heating. On the other hand, the electromagnetic energy is dissipating in the central region and gaining in the surrounding region. This suggests that in the electron diffusion region, both electromagnetic energy and internal energy are contributing to flow speed and outside the flow speed is approximately equally partitioned to electromagnetic and internal energy.

\section{\label{sec:phan}Electron-only Reconnection}
In 2018, MMS observed reconnection events where only the electrons participate in the reconnection process, but the ions do not~\cite{Phan2018Nature}. Although electron-only reconnection was already well known from numerical simulations~\cite{ShayEA98}, the 2018 events~\cite{Phan2018Nature} were the first time that such events were observed in natural plasmas by spacecraft. These novel ``electron-only" reconnection are different from the more traditional ``ion-coupled" reconnection in many different aspects~\cite{Stawarz2019ApJL}. Therefore, as a final case study, we present the energy conversion diagnostics in an electron-only reconnection event detected by MMS. An
analysis of pressure-strain dissipation in electron-only reconnection events has
not been presented previously.  
\begin{figure}
\includegraphics[width=\linewidth]{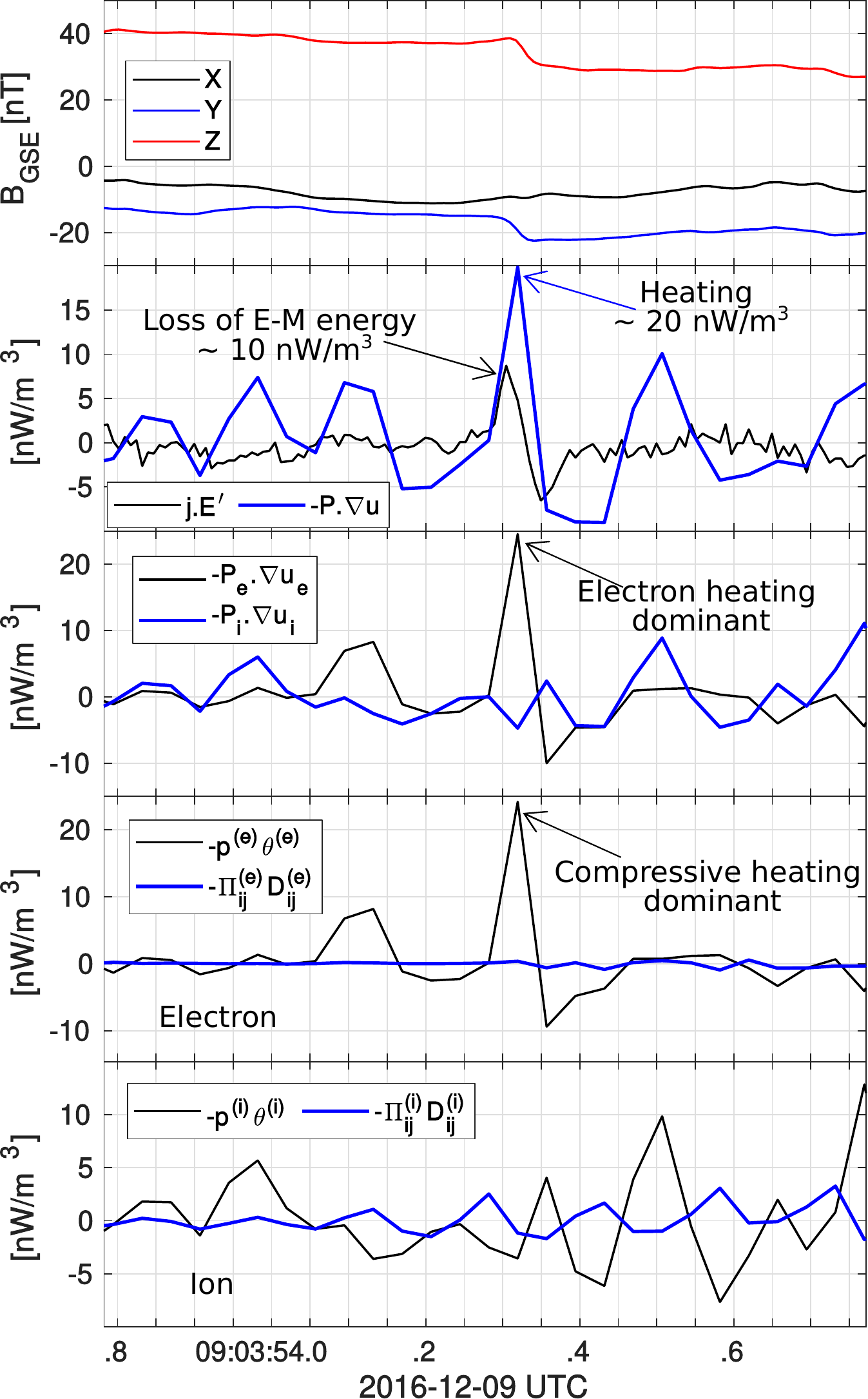}
\caption{\label{fig:phan}Electron-only reconnection in the magnetosheath, found by \citet{Phan2018Nature}.}
\end{figure}

In Fig.~\ref{fig:phan}, we show analysis of the ion and electron pressure-strain dissipation at an electron-only reconnection site,
observed by MMS. Both $\jde$ and $-\mathbf{P} \mathbf{\cdot} \mathbf{\nabla} \mathbf{u}$ are positive. Similar to the previous examples of magnetopause and magnetosheath reconnections, the magnitude of electromagnetic energy dissipation rate and the rate of change of internal energy are close to each other, but not exactly equal. The internal energy increase rate is somewhat higher than the electromagnetic energy dissipation rate. This observation supports the idea that increase of internal energy happens via flow energy and not directly from the electromagnetic energy. During this event, likely the flow is losing more energy than is being pumped in by the electromagnetic interactions.

The present analysis shows that this occurs through the intermediary of electron 
shear flows that couple through the pressure tensor directly into heat. 
In the previous samples we found that electrons contribute majority of the rate of change in internal energy. However, here we see that the ions do not show any enhancement at the reconnection site, suggesting their non-participation. The electron contribution alone accounts for nearly the total dissipation at the reconnection site. Nevertheless, we 
see that most of the energy
conversion is being facilitated through
the compressive channel, as we found in the other reconnection cases. 


\section{PIC Simulation}\label{sec:pic}
To further examine the nature of dissipation in turbulent reconnection sites, we perform fully kinetic turbulent PIC simulations of a fully ionized proton–electron plasma using the  P3D~\cite{Zeiler2002JGR} code. We perform the simulations in a 2.5D setup, which means that all field vectors have three components, but they vary only in a two-dimensional spatial grid $(X,Y)$ and there is no variation in the $Z$ direction.

All the results are in 
dimensionless 
units in which 
number density is normalized to a reference number density $n_r$, mass to ion mass $m_i$, charge to ion charge $q_i$, and magnetic field to a reference value $B_r$. Length is normalized to the ion inertial length $d_i$, time to the ion cyclotron time $\omega_{ci}^{-1}$, velocity to a reference Alfv\'en speed $v_{A} = B_r / \sqrt{4 \pi m_i n_r}$, and temperature to $T_r = m_i v_{A}^2$. The simulation was performed in 
a square periodic spatial domain 
represented by $4096^2$ grid points and initialized with 3200 
(macro-) particles of each species per cell. The length of each side is $L = 149.5648\,d_i$, so the grid size is $\Delta X = 0.037\,d_i \approx \lambda_D$, where $\lambda_D$ is the Debye length. The time step is $\Delta t = 0.005\,\omega_{ci}^{-1}$. We use an ion to electron mass ratio of $m_i/m_e = 25$ and a ratio of speed of light to Alfv\'en speed of $c/v_{A} = 15$ in this run. 

The simulation represents a decaying turbulence system, starting with uniform density $(n_0 = 1.0)$ and temperature of ions and electrons $(T_0 = 0.3)$. The mean magnetic field is $B_0 = 1.0$ directed out of the plane of the simulation box. The ion and electron beta (ratio of thermal to magnetic pressure) values 
are $\beta_{i} = 0.6$, $\beta_e = 0.6$, respectively.

\begin{figure}
\includegraphics[width=0.95\linewidth]{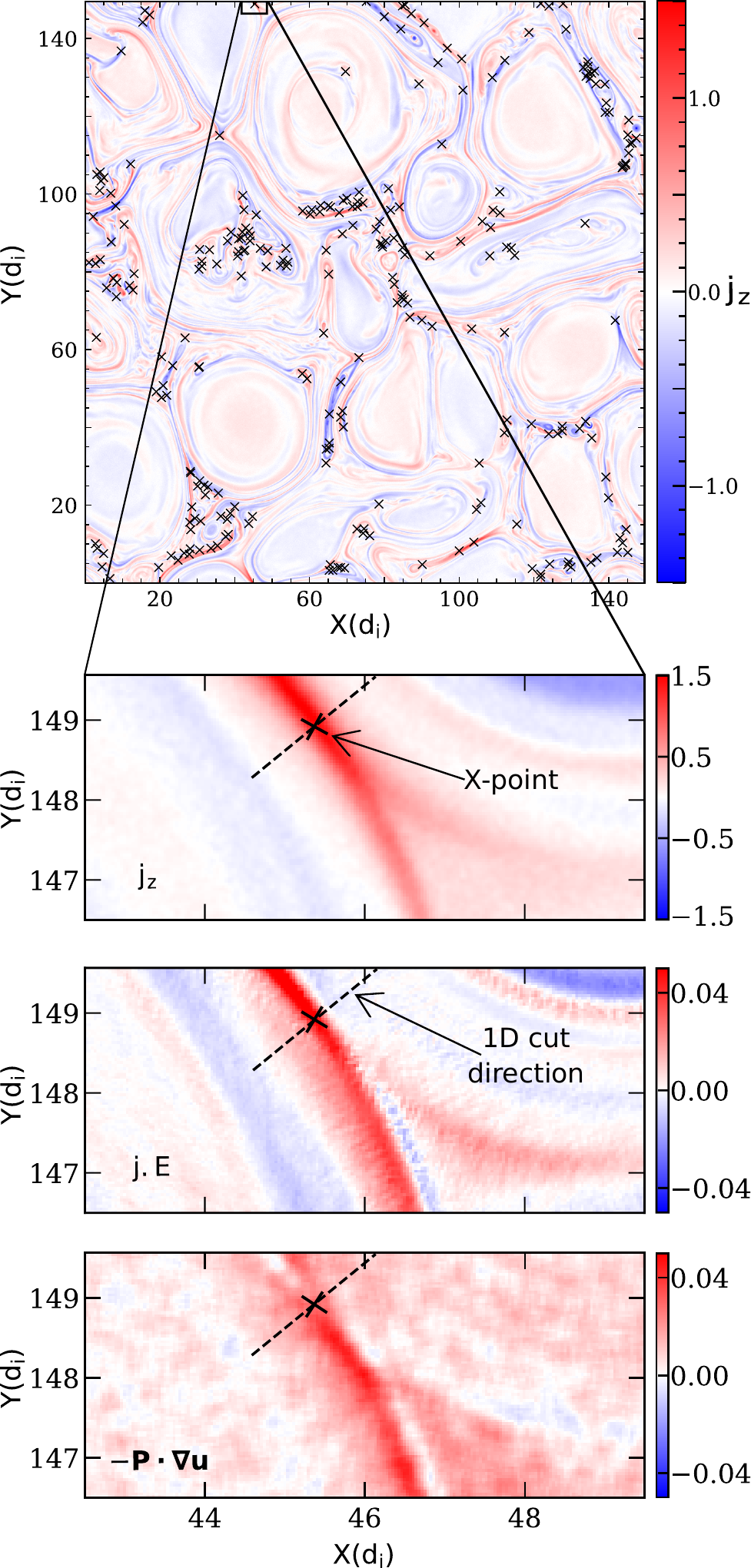}
\caption{\label{fig:2dpic}First panel: Color plot of out-of plane current ($j_{\mathrm{z}}$) together with X points (black crosses). Second: Enlarged view of the out-of plane current; Third: electromagnetic energy conversion rate; Fourth: the total pressure-strain around a chosen X point. The direction of the one-dimensional trajectory through the X point, shown in more detail in Fig.~\ref{fig:1dcut}, is indicated by the dashed lines.}
\end{figure}

We perform a case study using a snapshot after the time of maximum root mean square current density ($220\,\omega_{ci}^{-1}$). The system is in highly non-linear state at this point, well past the initial transient phase; yet, the turbulent fluctuation amplitudes have not decayed to very low values (see \citet{Parashar2018ApJL} for context and more details). To achieve sufficient statistical smoothness, we reran the PIC code for $0.6\,\omega_{ci}^{-1}$ starting from time $220\,\omega_{ci}^{-1}$, and save the full outputs at every timestep (i.e., $0.005\,\omega_{ci}^{-1}$). We average all the quantities over the $120$ time steps. Further, we perform a spatial smoothing by averaging over 3 neighboring points in each direction. The method described in \citet{Haggerty2017PoP_Xpt_statistics} (also see \onlinecite{Servidio2009PRL, Servidio2010PoP}) is used to find reconnection X-points. There are some subtleties associated with the number of X-points in numerical simulations, in that one needs also to understand whether the identified X-points are physical~\cite{Comisso2016PoP_general_theory}, or if they are numerical artifacts~\cite{Matthaeus1985PoF, Wan2010bPoP}. \citet{Haggerty2017PoP_Xpt_statistics} performed a careful analysis and found that a Fourier filter applied at the Debye scale is an optimal choice for eliminating the possible effects of numerical noise in identification of X-points in PIC data, and here we follow the same methodology. After performing these procedures, we find 273 critical saddle points. The top panel in Fig.~\ref{fig:2dpic} shows the out-of plane current in the whole simulation box and the X-points identified within. Most X-points are clustered around strong current sheets near magnetic island boundaries. 

\begin{figure}
\includegraphics[width=\linewidth]{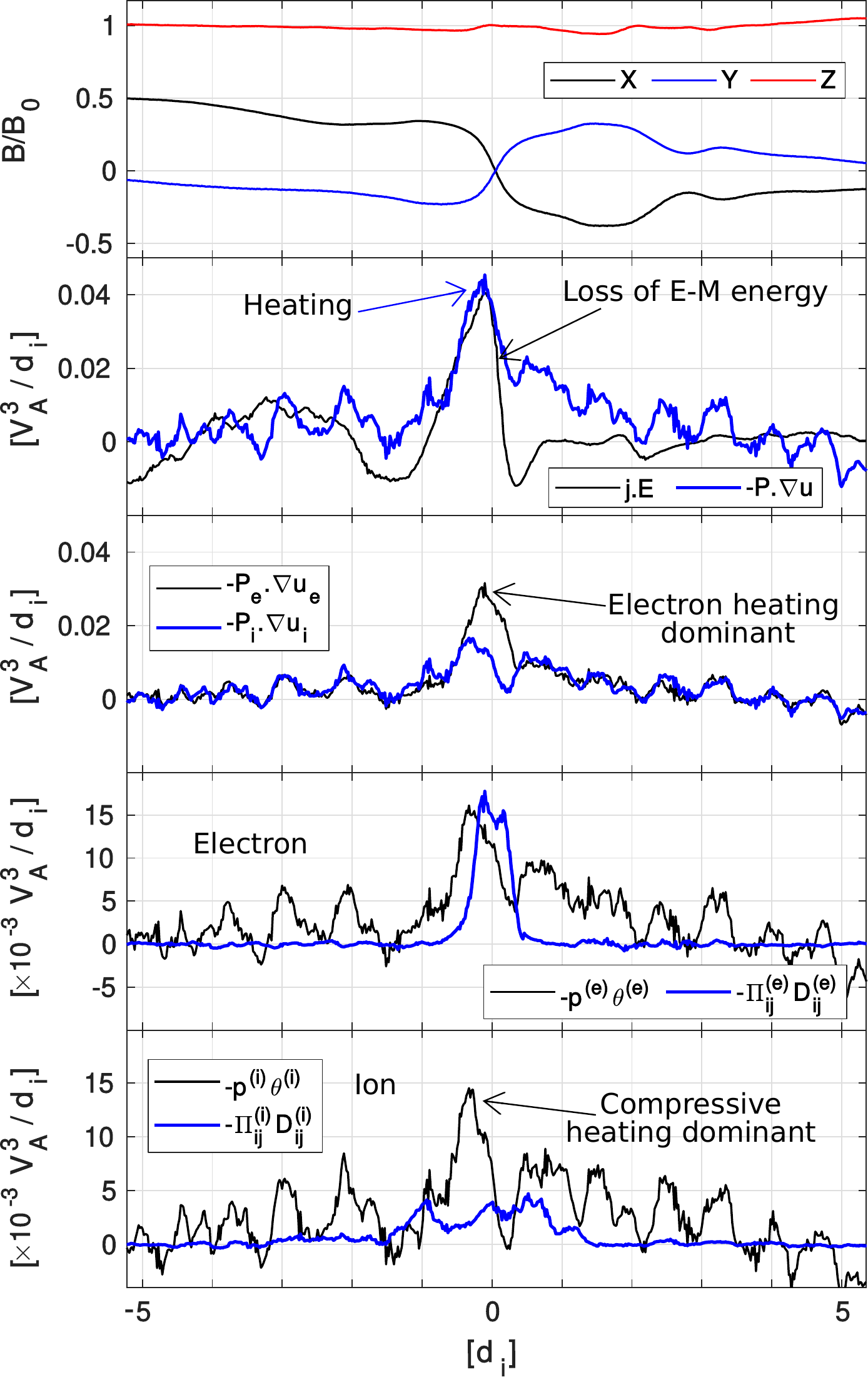}
\caption{\label{fig:1dcut} Magnetic field components and energy dissipation measures along the 1 dimensional cut through the X point in a 2.5D PIC simulation, shown in the right panels of Fig.~\ref{fig:2dpic}.}
\end{figure}

We focus on a particular X-point with strong current, at position $(45.37\,d_{i}, 148.92\,d_{i})$, denoted by a black rectangle in the top panel of Fig.~\ref{fig:2dpic}. The particular X-point is chosen because it has a high value of current, and the X-point is distinctly positioned in the current sheet between two magnetic islands with no other X-points nearby~\footnote{We performed the same analysis on 2 other strong current X points and found essentially the same behavior (not shown).}. The bottom three panels show an enlarged view of the region surrounding the X-point. The second panel clearly shows current enhancement near the X-point. The third panel shows enhancement of electromagnetic energy dissipation rate, and the bottom panel shows enhancement of internal energy conversion rate near the X-point. To further investigate the energy conversion channels into charged species, we record the measurements along a one dimensional sample through the X point. The direction of this sample is shown with dashed lines in the right panels of Fig.~\ref{fig:2dpic}. The magnetic field and energy conversion measures along this trajectory are shown in Fig.~\ref{fig:1dcut}.

From Fig.~\ref{fig:1dcut}, we see very similar characteristics as the MMS observations presented in the previous sections. First, from the second panel we see that the charged particles are heating and electromagnetic energy is being depleted. This behavior is consistent with the scenario of dissipating all particle energy, gained from electromagnetic fields, into heat. The third panel shows that electron heating rate is much greater than the ion heating rate, as seen in most MMS cases. Notably, this X-point has a high guide field. Unlike the MMS high-guide field, the electrons are not cooling here. This observation supports our previous suggestions that the cooling of the electron is probably controlled by some other factors and not only the guide field strength. The plots in the fourth panel indicates both compressive and incompressive channels of heating are effective for the electrons. However, the ions are heating mostly through the compressive channel, as shown in the last panel. This result contrasts with all the previously shown MMS results, which feature a large contribution from the compressive electron heating (or cooling) component, while ions are not significantly heated via the compressive term. The artificial electron mass used in the simulation could be the source of this discrepancy; however, a larger statistical study would be required to reach a more definite conclusion.

\section{Conclusions and Discussion}
Previous observational studies, performed using MMS data in the Earth’s magnetosheath, have focused on individual current sheets~\citep{Chasapis2018ApJ} or global turbulence statistics~\citep{Bandyopadhyay2020bPRL}. However, the behavior of pressure-strain interaction in reconnection sites, a crucial link in production of internal energy, has not been studied in detail, to the best of our knowledge. We note here that other approaches, such as tracking the net energy transfer between turbulent fields and plasma particles in velocity space with the field-particle correlation technique~\cite{Klein2016ApJL, howes2017JPP_FP_correlation} has the advantage of detecting a specific dissipation mechanism. The pressure-strain-based measures, on the other hand, are agnostic of dissipation mechanisms. However, the pressure strain based measures can differentiate between the transfer of energy between fields and fluid flow, and between fluid-flow and internal energy.
In this paper,  
we demonstrate the 
use of 
$\bf{J} \cdot {\bf E}$,
$- p \theta$,
and $\pid$ as independent 
observational diagnostics in reconnection using several case studies. 

Selected cases of different types of reconnection are presented. We find that the pressure-strain interaction can be used as an independent diagnostic of particle energization at reconnection sites. Although electromagnetic energy dissipation rate is positive in all the reconnection sites, pressure-strain shows particles can be either losing or gaining internal energy at the reconnection regions. However, in all cases the magnitudes of both measure are not very different. The rate of change of internal energy is larger for the electrons compared to the ions. This stronger electron energization may be due to the lower mass of the electrons. It is possible that the electrons, being lighter in mass, respond to the reconnection process quickly and get energized faster than the ions. Interestingly, simulations of kinetic plasma turbulence~\cite{Parashar2016ApJ_prop,Matthaeus2016ApJL,Yang2018MNRAS} as well as observations in the magnetosheath~\cite{Bandyopadhyay2020bPRL} 
and solar wind~\cite{Cranmer2009apj}
frequently report a higher {\it global} heating rate of ions compared to that of electrons. 
An explanation for this disparity remains elusive at present and is deferred to future work. 

We also find that the compressive channel of energy conversion is much stronger than the incompressive channel overall, and most of the time individually for each charged species. This observation is consistent with a recent result, obtained in a  current sheet within a turbulent simulation~\cite{Pezzi2021MNRAS_dissipation}. This behavior was not observed in `stand-alone' spontaneous 
reconnection simulations, and the compressive channel can evidently become stronger in a turbulent environment. Therefore, we here adopt the explanation offered by \citet{Pezzi2021MNRAS_dissipation} - the reconnection events occurring in thin current sheets, are often at the boundaries of magnetic islands. The magnetic islands compressing against each other due to external driving by turbulent motions might be the reason the compressive channel is stronger. Another factor could be due to the dependence of the plasma compressibility with the plasma beta. Larger plasma beta indicates a more compressible plasma. However, the MMS observations are all in the magnetosheath plasma, which have high beta. Therefore, since the larger compressive heating/cooling is observed, at least for ions, in both low beta (simulation) and high beta (MMS) systems, the role of plasma beta is not clear here. There are several control parameters, such as the plasma beta, guide-field strength, artificial mass of electrons, upstream parameters that may influence the ion to electron heating rate as well as the relative magnitude of compressive versus incompressive heating rates. With these limited number of case studies that we present here, it is not possible to scrutinize these possibilities.

Although the pressure-strain rate is weaker for ions compared to electrons in the reconnection cases studied, 
the electron-only reconnection case shows no signals in the ions at all. So, electron-only reconnection may be  completely due to electron energization without any participation from the ions, unlike the traditional reconnection, where the ion response is small, but  non-vanishing.

This paper presents case studies selected from prior analyses that did not employ the pressure strain diagnostics. The MMS mission has now observed many more large-scale magnetopause reconnection, as well as small-scale magnetosheath reconnection X-lines. A detailed survey with a larger number of samples will reveal the statistical nature of ion and electron energization and their relationship with electromagnetic field dissipation.

\begin{acknowledgments}
We wish to acknowledge NASA Heliospheric GI grant 80NSSC21K0739
at Princeton and support from the MMS project,
NASA NNX14AC39 under Theory and Modeling team grant 
80NSSC19K0565 at Delaware.

\end{acknowledgments}

\section*{Data Availability}\label{sec:data}
All MMS data are available at \href{https://lasp.colorado.edu/mms/sdc/}{https://lasp.colorado.edu/mms/sdc/}.

%

\end{document}